\documentclass[aip,
reprint]{revtex4-1}

\usepackage{graphicx}
\usepackage{dcolumn}
\usepackage{bm}
\usepackage[utf8]{inputenc}
\usepackage[T1]{fontenc}
\usepackage{mathptmx}
\usepackage{amsmath}
\usepackage{float}
\usepackage{amssymb}
\usepackage{color}
\begin{document}
	\preprint{AIP/123-QED}
	\title{Adsorption time scales of cluster-forming systems}

	\author{E. Bildanau}
	\affiliation{Belarusian State Technological University, 220006 Minsk, Belarus}
	
	\author{V. Vikhrenko}
	\affiliation{Belarusian State Technological University, 220006 Minsk, Belarus}
\begin{abstract}
A microscopic model of adsorption in cluster forming systems with competing interaction is considered. The adsorption process is described by the master equation and modelled by a kinetic Monte Carlo method. The evolution of the particle concentration and interaction energy during the adsorption of particles on a plane triangular lattice is investigated. The simulation results show a diverse behavior of the system time evolution depending on the temperature and chemical potential and finally on the formation of clusters in the system. The characteristic relaxation times of adsorption vary in several orders of magnitude depending on the thermodynamic parameters of the final equilibrium state of the adsorbate. A very fast adsorption of particles is observed for highly ordered adsorbate equilibrium states.
\end{abstract}

\maketitle
\section{Introduction}
\label{sec:intro}
Currently, there is a high activity in the study of the processes of self-organization and self-assembly in various systems. The elements of such systems are supramolecular formations with a molecular mass from units to thousands of kDa that lead to low rates of their thermal motion and sufficiently large, on molecular scales, characteristic times of the processes in them. Examples of such systems are solutions of protein molecules \cite{cardinaux:11:0,gunton:07:0,hunter:01:0}, which are of great interest in the biological and medical aspects; colloid metal or semiconducting nanoparticles and various types of core-shell particles that find numerous applications in  catalysis, optics, smart materials, etc. \cite{isa:17:0,vasudevan:18:0,sheverdin:19:0}; clays and soil suspensions \cite{bergaya:13:0}, widely used in industrial construction and agriculture; and many others.

Monolayers of macromolecules on gas-liquid, liquid-liquid or liquid-solid interfaces are of great importance due to their ability to stabilize emulsions and foams, to form self-assembled two-dimensional (2D) ordered structures that can find applications in plasmonic systems, anti-reflecting coating, for sensing, etc. \cite{dimitrov:96:0,chattopadhyay:10:0,yao:14:0} There exists a vast variety of experimental investigations of adsorption processes concerned with the deposition of particles on surfaces and interfaces \cite{graham:79:0,beverung:99:0,wertz:99:0,schwartz:01:0,marie:02:0,schreiber:04:0,rabe:11:0,vogler:12:0,langdon:12:0,hasan:16:0}. The significance of cooperative effects in the adsorption processes is frequently emphasised \cite{rabe:11:0,nygren:94:0,ball:97:0,herring:06:0,vanderveen:07:0,rabe:08:0,rabe:10:0}.

At the same time, the interaction between these elements is very complex, and, despite their rather large sizes compared to molecular ones, the scale of the interparticle interaction energy may remain insignificantly larger of the thermal energy of the order of several $k_BT$  in the range of approximately room temperatures\cite{cardinaux:07:0,cardinaux:11:0,zhuang:16:0,royall:18:0}. This provides rich opportunities for various phase transitions in such systems at room temperatures because it is well known from the theory of liquid state\cite{huang:87:0,barker:76:0,hansen:06:0} that critical temperatures are close to the pit depth of the interparticle interaction potential divided by the Boltzmann constant. It has been established \cite{seul:95:0,sear:99:0} that in many cases the formation of cluster phases is a result of competing interparticle interactions, e.g. van der Waals attraction at short and Coulomb repulsion at larger distances (SALR systems -- Short range Attraction Long rang Repulsion). 

The need to understand the processes occurring in the systems of the types described above, the possibility of predicting their behavior in various conditions and controlling their properties requires the development of statistical-mechanical methods for their study. In principle, the methods for studying molecular systems are well developed \cite{barker:76:0,huang:87:0,hansen:90:0}, however, the large masses of particles and the peculiarity of interparticle interactions lead to the need for a substantial modification of the developed methods. Conventional liquid state theories describing their thermodynamic and structural properies as well as non-equilibrium behavior are based on the binary distribution functions and integral equations for them, while the systems with competing interparticle  interactions are characterized by existence of clusters containing many particles that requires many-particle distribution functions for their characterization. The distribution of particles among the clusters of different sizes and the distribution of clusters over the system volume plays an important role in the behavior of such systems\cite{stradner:04:0,zhuang:16:2,santos:17:0,royall:18:0,litniewski:19:0,bildanau:20:0}. 

2D lattice models of the systems with competing interactions are widely studied due to possibility understanding many their fundamental features with comparatively restricted computational facilities. To date, main efforts were concentrated on investigating microphase separation and pattern formation in bulk \cite{imperio:06:0,schwanzer:10:0,pekalski:14:0,almarza:14:0,chacko:15:0,das:15:0} and confined \cite{archer:08:0,almarza:16:0,pekalski:19:0,bildanau:20:0} equilibrium systems. Kinetic properties were rarely addressed. In Ref. \cite{schwanzer:16:0} scattering functions and diffusion properties of individual particles and clusters in an equilibrium SALR system were considered. The clustering dynamics in 1D systems was considered in \cite{hu:18:0}. Some attempts were undertaken to model the protein adsorption based on microscopic representations \cite{rabe:10:0,zhou:03:0,pellenc:08:0,yu:14:0}.  

Adsorption is a complicated process that can be controlled by many factors such as diffusion in the bulk solution, barrier resistance, processes in the near surface layer, reorientation and conformation changes of adsorbed particles, and so on\cite{graham:79:0,beverung:99:0,wertz:99:0,schwartz:01:0,marie:02:0,schreiber:04:0,rabe:11:0,vogler:12:0,langdon:12:0,hasan:16:0,nygren:94:0,ball:97:0,herring:06:0,vanderveen:07:0,rabe:08:0,rabe:10:0}. In the present contribution, we investigate the influence of competing interactions on the kinetics of adsorption from a solution neglecting the lateral diffusion. A main attention is paid to the time scales of the process and manifestation of cooperative effects attributed to interparticle interactions and formation of cluster structures on the interface.

\section{Model}
\label{sec:model}

To study the kinetics of adsorption of the system with interparticle competing interaction, we consider deposition of particles on a flat surface from a fluid (gas or liquid) phase where the state of the particles is characterized by their chemical potential $\mu^*$. The dynamics of the system on the surface is carried out through the processes of adsorption and desorption of particles starting from the vacuum state and without accounting of diffusion on the surface. 

The particles on the surface are characterized by the chemical potential $\mu^*$ and lateral interparticle interactions. As in Refs~\cite{pekalski:14:0,almarza:14:0,pekalski:15:0,almarza:16:0,pekalski:19:0,bildanau:20:0}, the lattice model on a close packing triangular lattice is considered. 
%The surface contribution to the chemical potential can be determined by the interaction with the surface and surface pressure. 
The periodic boundary conditions and a system size $L \times L = 60 \times 60$ are used to minimize the confined effects in Monte Carlo (MC) simulation. To understand the size effect some simulations of the system with $L=120$ were performed. The average values of the required quantities (particle concentration, system energy) as functions of time are determined by averaging over about 20 thousand trajectories to get a more statistically reliable results. The longest trajectories were of 8 000 Monte Carlo steps (MCS) during which the concentration definitely reached the equilibrium value.

The SALR interaction potential between the particles on the surface is taken in accordance with Refs.~\cite{almarza:14:0,almarza:16:0}:

\begin{equation}
	V^*(\Delta \mathbf{x}) = 
	\begin{cases}
		-J_1 \quad \textrm{for $|\Delta \mathbf{x}| = 1,$ \quad  for nearest neighbors} \\
		+J^*_3 \quad \textrm{for $|\Delta \mathbf{x}| = 2,$ \quad  for third neighbors} \\
		0 \qquad \textrm{otherwise,} 	
	\end{cases} 
\end{equation}
where $-J_1$ and $J^*_3=J_3J_1$ represent the energy of interparticle attraction and repulsion, respectively, $\mathbf{x}$ is the radius-vector of a lattice site, $|\Delta \mathbf{x}|$ is the distance between particles on the corresponding lattice sites. The ratio $J_3 = J^*_3/J_1 = 3$ is used as in Refs. \cite{pekalski:14:0,almarza:14:0,almarza:16:0}.

The thermodynamic Hamiltonian of the system is as follows:
\begin{equation}
	H = \frac{1}{2} \sum_{\mathbf{x}}\sum_{\mathbf{x'}}\hat{\rho}(\mathbf{x})V(\mathbf{x}-\mathbf{x'})\hat{\rho}(\mathbf{x'}) - \mu\sum_{\mathbf{x}}\hat{\rho}(\mathbf{x}),
	\label{eq:hamiltonian}	
\end{equation}
where $\sum_{\mathbf{x}}$ is the sum over all lattice sites, $\hat{\rho}(\mathbf{x})$ is the occupation number. $\hat{\rho}(\mathbf{x})=1$ or 0 if the site with the coordinate $\mathbf{x}$ is occupied or vacant. In simulations, the dimensionless values of the interparticle interaction energy $V=V^*/J_1$, temperature $T=k_B T^*/J_1$ and chemical potential $\mu=\mu^*/J_1$ are used, $k_B$ is the Boltzmann constant. In our simulation, we used temperatures below critical $T = 0.80$, around critical $T = 0.95$ and above critical $T = 1.20$. The phase diagram of the system was built\cite{almarza:14:0} at $L=120$. In our case, the system was smaller ($L=60$), and at periodic boundary conditions the characteristic temperatures can be slightly larger\cite{landau:76:1,landau:83:0}. 

The time evolution of such systems is usually described by the master equation\cite{mcquarrie:67:0,binder:74:0,vankampen:81:0}
	\begin{equation}
		\begin{split}
			{\frac {P(\{\mathbf{x}\},t^*)}{dt^*}}&=\sum_{\{\mathbf{x'}\} \neq \{\mathbf{x}\}}[W(\{\mathbf{x'}\}\rightarrow \{\mathbf{x}\})P(\{\mathbf{x'}\},t^*) \\ &-W(\{\mathbf{x}\}\rightarrow \{\mathbf{x'}\})P(\{\mathbf{x}\},t^*)],
		\end{split}
		\label{eq:master}
\end{equation}
where $P(\{\mathbf{x}\},t^*)$ is the probability for a particular distribution $\{\mathbf{x}\}$ of particles over the lattice sites, $W(\{\mathbf{x}\}\rightarrow \{\mathbf{x'}\})$ is the transition rate for the particle distribution change from $\{\mathbf{x}\}$ to $\{\mathbf{x'}\}$, which has to obey the detailed balance condition.

The master equation describes the behavior of the ensemble of systems at given initial and external conditions. This behavior can be approximated by MC simulation of a large number of identical systems at these initial and external conditions. The behavior of each system of the ensemble is described by the microscopic counterpart of the master equation\cite{danani:97:0,bokun:01:0,alanissila:02:0}. For the case of adsorption/desorption process it can be written as
	\begin{equation}
		{\frac {d\hat \rho(\mathbf{x},t^*)}{dt^*}}=-{\hat \rho(\mathbf{x},t^*)}W_d(\mathbf{x},t^*)+(1-{\hat \rho(\mathbf{x},t^*)})W_a(\mathbf{x},t^*),  
		\label{eq:dynamics}
	\end{equation}

where the thermally activated rates of the particle desorption or adsorption is correspondingly determined by the expression
\begin{equation}  
	W_d(\mathbf{x},t^*)=
	\begin{cases}
		\nu_d\exp[H_1(\mathbf{x},t^*)/T] \quad & \textrm{for $H_1(\mathbf{x},t^*) \leq 0$,} \\
		\nu_d \qquad \qquad  & \textrm {for $H_1(\mathbf{x},t^*)>0$},
	\end{cases} 
	\label{eq:probab_des}
\end{equation}
or
\begin{equation}
	W_a(\mathbf{x},t^*)=
	\begin{cases}
		\nu_a\exp[-H_1(\mathbf{x},t^*)/T] \quad & \textrm{for $H_1(\mathbf{x},t^*) > 0$,} \\
		\nu_a \quad & \textrm {for $H_1(\mathbf{x},t^*)\leq 0$},
	\end{cases} 
	\label{eq:probab_ads}
\end{equation}
where 
\begin{equation}
	\begin{split}
		H_1(\mathbf{x},t^*)&=V_1(\mathbf{x},t^*)-\mu \\
		V_1(\mathbf{x},t^*)&=\sum_{\mathbf{x'}}V(\mathbf{x}-\mathbf{x'})\hat{\rho}(\mathbf{x'})
	\end{split}
	\label{eq:energy}
\end{equation}
is the particle ($H_1$) or particle interaction ($V_1$) energy on the surface, $\nu_d$ and $\nu_a$ are the frequency prefactors\cite{alanissila:02:0,jansen:12:0} that can be evaluated in the framework of the transition state theory or considering the particle dynamics in the surface adjoining layer of the solution. The prefactors determine the time scales of the adsorption process. 

Eq.\ref{eq:dynamics} sometimes is also referred to as the master equation and the generalized time derivative\cite{dirac:58:0,glegg:17:0} in it determines the microscopic fluxes. Time in MC simulation is a discrete quantity with the time step $\Delta t^*$ equal to one trial to change the system state. In this case the time derivative has to be replaced by the ratio $\Delta \hat{\rho}(\mathbf{x},t^*)/\Delta t^*$ with $\Delta \hat{\rho}(\mathbf{x},t^*)=\pm 1$ or 0 depending on the probability realization.

In fact, Eq.(\ref{eq:probab_ads}) models the sticking probability~\cite{jung:00:0} because larger the energy of adsorption $H_1(\mathbf{x},t^*)$ smaller the probability of adsorption. 

In MC simulation, time is usually measured in Monte Carlo steps (MCS consists of one trial per particle). Kinetic Monte Carlo methods\cite{jansen:12:0} provide with a number of algorithms for transferring MCS into physical time. On the other hand, the master equation allows to use the frequency prefactors for transferring MCS into physical time. 

This conclusion can be supported by considering the particle diffusion on a lattice. In Monte Carlo simulation, the tracer diffusion coefficient is calculated in units of the diffusion coefficient at the limit of zero coverage. The time unit of the latter is the inverse prefactor frequency $\nu^{-1}$ that determines the time scale of the process~\cite{gomer:90:0}. E.g., the mean square displacement of a particle on a square lattice with the lattice parameter $a$ is equal~\cite{gomer:90:0,sadiq:83:0,uebing:91:0,bokun:01:0} $<(\Delta {\bf x})^2>=4D_{tr}D^*_0t_{MCS}$, where $D_{tr}$ is the tracer diffusion coefficient calculated through the Monte Carlo simulation, $D^*_0=(1/4) a^2\nu$ is the diffusion coefficient at the limit of zero coverage. $D_{tr}$ is the ratio of the physical diffusion coefficient $D^*_{tr}$ to the diffusion coefficient at the limit of zero coverage in a real physical system $D^*_{0}$ or the ratio $D_{tr}/D_0$, where $D_0=1$ is the diffusion coefficient at the limit of zero coverage in the lattice system. The tracer diffusion coefficient in the physical system is $D^*_{tr}=D_{tr}D^*_0$ and it follows from the expression $<(\Delta {\bf x})^2>=4D^*_{tr}t^*$ that $t^*=t\nu$ with $t=t_{MCS}$.

The remarkable feature of the transition rates Eqs.(\ref{eq:probab_des}), (\ref{eq:probab_ads}) is that at $\nu_a=\nu_d=\nu$ they can be transformed to the Metropolis' importance sampling algorithm~\cite {metropolis:53:0,landau:76:1,ito:93:0} satisfying the detailed balance condition \cite{binder:74:0,sekimoto:10}. To this end, Eq.(\ref{eq:dynamics}) has to be divided by $\nu$ and dimensionless time $t=\nu t^*$ has to be used where $t$ is measured in MCS. In this case, the final equilibrium state of the system corresponds to the Hamiltonian Eq.(\ref{eq:hamiltonian}) that has already been thoroughly investigated\cite{almarza:14:0} by the grand canonical MC simulation and the inverse value of the frequency prefactor $\nu^{-1}$ has to be used transferring from MCS to physical time: $t^*=t/\nu=\rm{MCS}/\nu$.   

With accounting of Eqs.(\ref{eq:probab_des}) and (\ref{eq:probab_ads}), Eq.(\ref{eq:dynamics}) after averaging over the non-equilibrium distribution can be written as
\begin{equation}
	\frac{dc}{dt}=
	\begin{cases}
		-c{\rm exp} \left[\frac{E(t)-\mu}{T} \right]+(1-c) \quad & {\rm for} \quad  E(t)-\mu \leq 0, \\
		-c+(1-c){\rm exp} \left[\frac{\mu-E(t)}{T} \right] \qquad  & {\rm for} \quad  E(t)-\mu>0,
	\end{cases} 
	\label{eq:kinetics}
\end{equation}
where $c(t)=\langle  \hat \rho(\mathbf{x},t) \rangle$ and $E$ are the mean concentration and desorption/adsorption activation energy of a particle in the reduced units, respectively. The latter at $E(t) \lessgtr \mu$ can be evaluated from the expression
\begin{equation}
	\begin{split}
		{\rm exp}\left[\frac {\pm E(t)}{T}\right]& =\frac {1}{cM}\bigg \langle \sum_{i=1}^{M}\hat {\rho_i}{\rm exp}\left[\frac {\pm V_1(\mathbf{x},t)}{T} \right] \bigg \rangle \\
		&=\frac {1}{cM}\bigg \langle \sum_{i=1}^{M}\hat {\rho_i}{\rm exp}\left[\frac {\pm (-z_{1i} J_1+z_{3i} J_3)}{T} \right] \bigg \rangle,
	\end{split}
	\label{eq:meanenergy}    
\end{equation}
where $\hat \rho_i \equiv \hat{\rho}(\mathbf{x},t)$ is the occupation number of site $i$, $z_{1i}$ and $z_{3i}$ are  the numbers of the first and third neighboring particles of a particle on the lattice site $i$ at time $t$, $M=L^2$ is the total number of lattice sites, the angular brackets mean the averaging over the non-equilibrium ensemble or over MC trajectories. 

\begin{figure}
	\centering
	\includegraphics[width=1\linewidth]{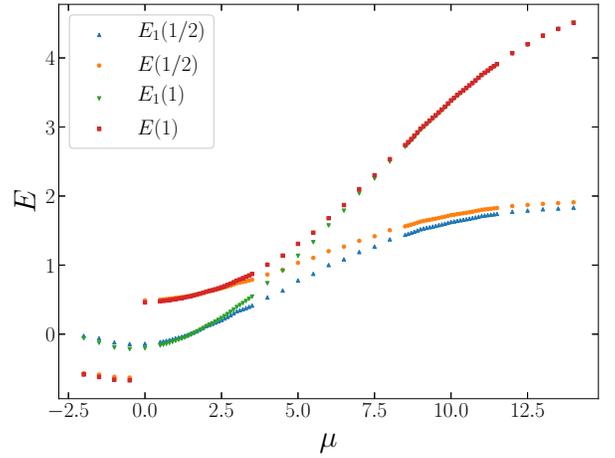}
	\caption{The average $E_1$ and mean desorption/adsorption activation $E$ energy of a particle on the interface at the half (1/2) and the end (1) of the first MCS at $T=0.8$ and different chemical potential values. }
	\label{fig:meanEnergy}
\end{figure}

The mean desorption/adsorption activation energy can be represented through the cumulant expansion. However, such a calculation is a very complicated task and a very crude estimation only can be done at some specific conditions. In Fig.\ref{fig:meanEnergy} the simulation results for the mean desorption/adsorption activation energy $E$ in accordance with Eq. (\ref{eq:meanenergy}) and the average energy $E_1$ of interaction of a particle with its surrounding (Eq.(\ref{eq:particlenergy})) are shown. 
\begin{equation}
	\begin{split}
		E_1(t) &=\frac {1}{cM}\bigg \langle \sum_{i=1}^{M}\hat {\rho_i} V_1(\mathbf{x},t) \bigg \rangle \\
		&=\frac {1}{cM} \bigg \langle \sum_{i=1}^{M}\hat {\rho_i}(-z_{1i} J_1+z_{3i} J_3) \bigg \rangle.
	\end{split}
	\label{eq:particlenergy}    
\end{equation}

The difference between the mean desorption/adsorption activation and average particle energy arises due to energy fluctuations because the  exponential function,which enters in Eq.(\ref{eq:meanenergy}) increases the value of the integral. The contribution of fluctuations decreases with increasing the average energy. 

At negative values of the chemical potential (small concentration) the mean energy is negative as well that manifests the influence of the nearest neighbor attractive interactions on the adsorption process. Repulsive interactions prevail at larger concentrations. 

\section{Results}
\label{sec:results}
	
\subsection{Evolution of the particle concentration}
\label{sec:conc}

The evolution of the system on the shortest time scale can be tracked by developing the first MCS into individual trials. Only at low values of the chemical potential when the interparticle interactions can be neglected $E_1\simeq 0$, it is possible to describe the process analytically. As it follows from Eq. (\ref{eq:kinetics}), the adsorption kinetics is of the first order (Langmuir non-cooperative type) and the concentration evolution is described by the expression (in reduced units)
\begin{equation}
c(t)=\left(1+{\rm exp}(-\mu/T)\right)^{-1}(1-{\rm exp}(-t/\tau)),
\label{eq:kinetics1}
\end{equation}
where the relaxation time is $\tau=[1+{\rm exp}(-|\mu|/T)]^{-1}$, which varies from 0.5 to 1 MCS. In real systems at these conditions the adsorption kinetics is diffusion limited\cite{graham:79:0} because requires fast supply of the adsorbed particles to the surface. At these conditions, the first time derivative of the concentration at $t=0$, $c=0$ in the MC simulation completely coincides with the analytical result $dc/dt={\rm exp}(\mu/T)$ at $\mu \leq 0$ or $dc/dt=1$ at $\mu>0$. This is an additional verification that $\nu^{-1}$ is the multiplier transforming the MCS into physical time.  

With an increase in the chemical potential and density of deposited particles, the interaction between the particles becomes important. As a result, even during the first MCS the concentration time dependence in the MC simulation strongly deviates from the analytical solution Eq.(\ref{eq:kinetics1}). Since $L^2$ trials are accomplished for one Monte Carlo step, the concentration during the first MCS rises to a rather high value dependent on the chemical potential. The deposited particles have to overcome the resistance of  particles already deposited on the surface, and the adsorption process becomes barrier limited\cite{graham:79:0}.  

\begin{figure}
	\centering
	\includegraphics[width=1\linewidth]{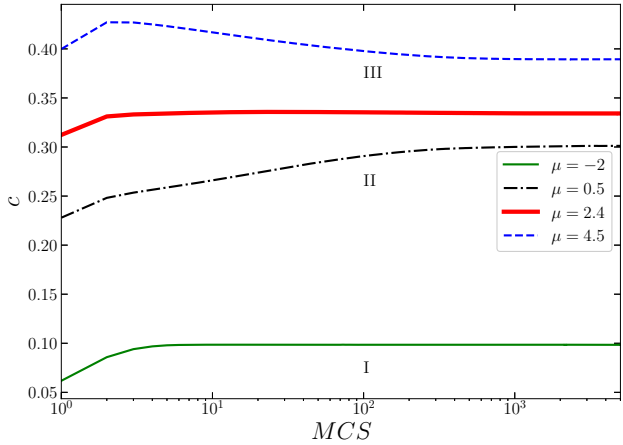}
	\caption{Three types of the concentration evolution at $T=0.8$ and different chemical potential values. The lower ($\mathrm{I}$), middle ($\mathrm{II}$) and upper ($\mathrm{III}$) curves represent the first, second and third types of the concentration time dependencies, respectively. A bold red line ($\mu=2.4$) indicate the relaxation curve with the high ordered state.}
	\label{fig:typesEvol}
\end{figure}

On the longer time scale, the adsorption intensity decreases and the dependencies of the mean particle concentration $c$ on the number of Monte Carlo steps are of three different types in different regions of the chemical potential (Fig. \ref{fig:typesEvol}) that can be characterised by the chemical potential values $\mu_F,\mu_R,\mu_L,\mu_B$, $\mu_D$ located in the regions of the disordered fluid, rhomboidal, fluid with lamella residues, bubbles (inverted rhomboidal) ordered phases and dense state with vacancies, respectively. They separate the regions of various types of the concentration time dependences. For three investigated isotherms $T=0.8, 0.95, 1.2$ they correspondingly are $\mu_F = -1.5$ $(c=0.14)$, $-1.0 (c=0.18)$, -0.5 $(c=0.22)$, $\mu_D=12.5$ $(c=0.78)$, $13.0 (c=0.82), 13.5 (c=0.86)$. $\mu_R=$2.4 $(c=1/3), \mu_L=6.0 (c=1/2), \mu_B=9.6 (c=2/3)$ at all the temperatures corresponding to highly ordered states of the equilibrium systems.  

\begin{figure}
	\centering
	\includegraphics[width=1\linewidth]{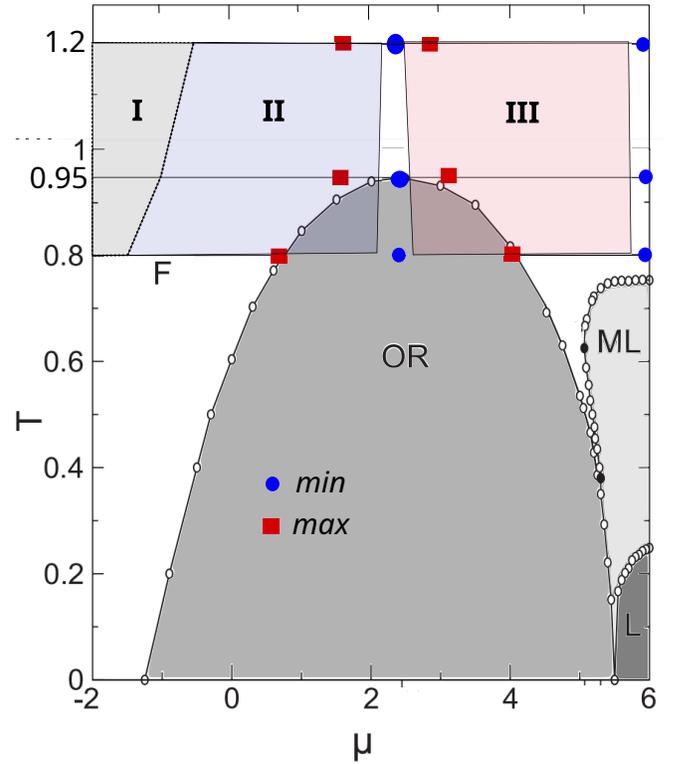}
	\caption{The disposition of the isotherms and points corresponding to the minimal (min) and maximal (max) values of the relaxation times on the phase diagram\cite{almarza:14:0} of the SALR system. F is a disordered fluid phase, OR and L are the ordered rhombus and lamella phases, respectively, ML is the molten lamella phase. Regions I, II, III determine the conditions under which the corresponding type of concentration relaxation is observed.}
	\label{fig:PhaseDiagram}
\end{figure}

For small $ \mu < \mu_F$ and large $\mu > \mu_D$ values of the chemical potential, the first simple type of reaching the equilibrium state can be described by an exponential function. For $\mu \in (\mu_F; \mu_{R}) \cup (\mu_L; \mu_B)$ and $\mu \in (\mu_R; \mu_L) \cup (\mu_B;\mu_D)$, the second and third types are observed, respectively. For the second type, the concentration time dependence is more complicated and cannot be described by a simple exponential function. The third type is characterised by a hump on the curve representing the concentration time dependence (the upper curve in Fig.~\ref{fig:typesEvol}) manifesting the overshooting effect\cite{rabe:08:0,rabe:11:0}. Physically, different types correspond to the formation of clusters in the system. The first type is characterized mainly by the presence of monomers and dimers, the second is a mixture of rhombuses of different orientations and triangular clusters, and the third is various fragments of stripes and clusters with more than 5 particles. The regions that determine the conditions under which the corresponding type of concentration relaxation is observed are shown in the phase diagram (Fig. ~\ref{fig:PhaseDiagram}).

For the chemical potentials that correspond to the third type of the concentration evolution, the excessive concentration of particles as compared to the equilibrium value is observed during the adsorption process. Within a relatively small number of Monte Carlo steps ($t \sim 10$ MCS), the particles are deposited to the surface without creating the short range ordering that corresponds to the equilibrium distribution. This type of concentration evolution corresponds to the chemical potentials when at low temperatures lamella ordering of the system is observed. At these conditions at earlier stage, the interparticle attraction plays more important role, while on the later stage the interparticle repulsion leads to establishing the equilibrium concentration and the final interparticle distribution. The overshooting effect appears in the system with spherical interparticle interaction in contrast to Refs.\cite{rabe:11:0,rabe:08:0} where this effect is explained by particle re-orientations on the surface.

In general, the adsorption process for these cluster-forming systems has a complicated nature and cannot be described by a few exponential functions. For the second and third types of evolution it was not possible to develop an identical fitting procedure based on exponential functions. Instead, the estimation of the total characteristic times of the adsorption depending on the chemical potential or equilibrium concentration was based on reaching the equilibrium concentration. The characteristic time can be estimated as time when the integral of the concentration deviation from its equilibrium value starts to be independent on the upper limit \cite{suzuki:71:0,binder:74:0,saito:79:0,kikuchi:93:0}. Averaging over 20 000 trajectories and additional smoothing over 11 MCS were used for calculating $\sum^{i_{\rm max}}_{i=1}|c_{\rm eq}-\sum^{5}_{j=-5}c_{i+j}/11|$, where the outer sum approximates the integral and equilibrium concentration $c_{\rm eq}$ was determined by averaging the concentration during last 100 MCS and over all 20 000 trajectories. It was checked that the sum converges to a constant value when the difference $|c_{\rm eq}-c_i|$ reaches the value $10^{-4}$. Then the total concentration relaxation time $\tau_{ct}=t_i$ was determined as time when the difference $|c_{\rm eq}-\sum^{5}_{j=-5}c_{i+j}/11|$ becomes equal or smaller of $10^{-4}$. 

\begin{figure*}
	\includegraphics[width=1\linewidth]{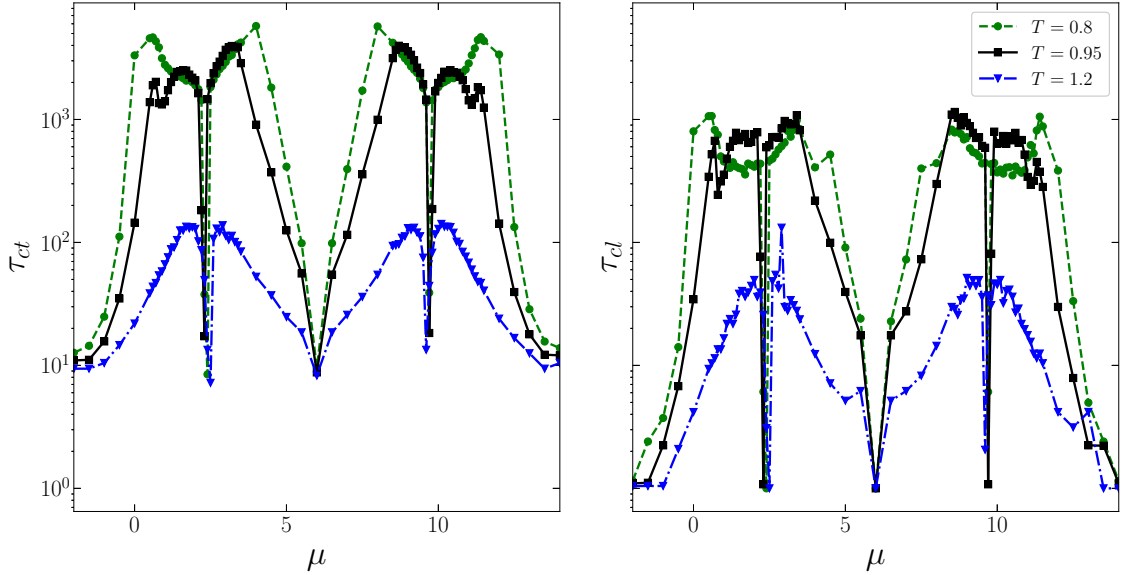}
	\caption{The total (left panel) and longest (right panel) concentration relaxation time for the SALR system $versus$ the chemical potential $\mu$. The deep minima are attained at concentrations $c=$1/3, 1/2 and 2/3 ($\mu=2.4$, 6.0 and 9.6, respectively) when highly ordered phases exist; see Fig.\ref{fig:DenseConc} representing the equilibrium concentration dependence on the chemical potential and Fig.\ref{fig:PhaseDiagram} showing the disposition of the ordered phases.} 
	\label{fig:TimeConc}
\end{figure*}

The obtained dependence of the concentration total relaxation time on the chemical potential is shown in Fig. \ref{fig:TimeConc} (left panel). The curves are symmetric with respect to $c=0.5 (\mu=6.0)$ due to the symmetry of the phase topology in the system: particles are replaced by vacancies and the phase of ordered rhombuses is replaced by the phase of ordered rhombus bubbles.

Additional characterization of the time evolution during equilibrisation consists of a sequence of relaxation times on different time intervals\cite{stoll:73:0}. Such a sequence can be determined as time intervals during which the deviation from the equilibrium value decreases by e times. In our case such relaxation times increase when the time increases. The longest of these relaxation times $\tau_{cl}$ can be estimated as the time interval between the moments when the concentration reaches values that differ from $c_{\rm eq}$ by ${\rm e}\cdot 10^{-4}$ and $10^{-4}$ that corresponds to the time dependence $|c_{\rm eq}-c(t)|\sim {\rm e}^{-t/\tau_{cl}}$ on this time interval. The simulated concentration time dependence was smoothed over 11 points as it was described in the previous paragraph. Because $\tau_{cl}$ characterises a part of the relaxation curve, it is shorter of the total relaxation time (Fig.\ref{fig:TimeConc}, right panel). 

\begin{figure}
	\centering
	\includegraphics[width=1\linewidth]{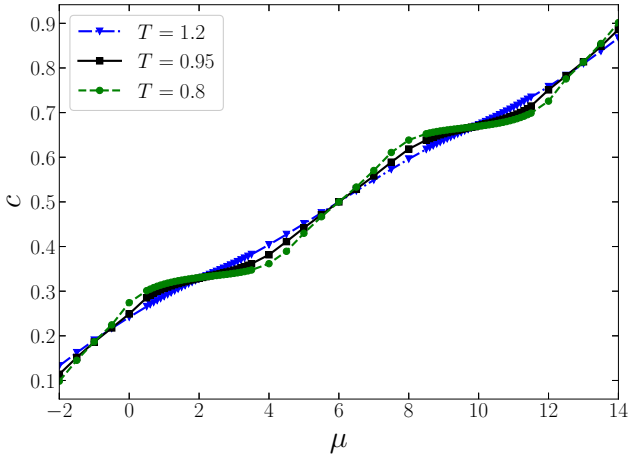}
	\caption{Dependence of the mean equilibrium concentration on the chemical potential at different temperatures. The isotherms intersect at the chemical potentials $\mu=2.4, 6.0, 9.6$ that correspond to highly ordered states at concentrations $c=1/3, 1/2, 2/3$.}
	\label{fig:DenseConc}
\end{figure}

The fastest achievement of concentration equilibrium is observed in the ordered region for the mean particle concentrations $c = 1/3, 1/2, 2/3$, which correspond to the chemical potentials $\mu=2.4,$ 6.0, 9.6, respectively. The concentration isotherms are shown in Fig.\ref{fig:DenseConc}. It is worth noting that in the ordered regions of rhombuses and bubbles, the relaxation curves of the second and third types merge (the curve for $\mu=2.4$ on Fig. ~\ref{fig:typesEvol}). It means that in the system, both parts of the interaction potential (attractive and repulsive) under the conditions of the existence of a highly ordered phase behave equally.

Comparing with the phase diagram (Fig. \ref{fig:PhaseDiagram}) for these systems, we can conclude that the points of maximal values of the relaxation time ($\mu = 0.6$ and $\mu=4.0$) for the temperature $T = 0.8$ correspond to the phase transition points from the disordered fluid state (F) to the phase of ordered rhombuses (OR) at $\mu<6.0$ or rhombus bubbles at $\mu>6.0$. These regions are characterized by slowdown of the adsorption process in the vicinity of the phase transition points in analogy with the critical slowdown of kinetic processes in the systems with simpler interactions\cite{hohenberg:77:0}. Due to not large enough the system size, the phase transition to the ordered rhombuses is not exactly of the first order. The concentration isotherms are smooth curves without empty regions corresponding to the phase coexistence (Fig.\ref{fig:DenseConc}). They mix the features characteristic of the critical point (an anomalous slowdown) like it was observed\cite{zhang:17:0} in the study of the dynamical behavior of a polymer grafted onto an adhesive surface. Moreover, in our study the system in the near critical region ($\mu=0.6$, 4.0 or 8.0) demonstrates suppressing the fluctuations and strong speeding up the relaxation. A noticeable acceleration of the relaxation of the system (minima of the $\tau_{ct}$ - $\mu$ curve) after the region of critical deceleration (maxima of the $\tau_{ct}$ - $\mu$ curve) is due to the presence of a highly ordered phase in which significant the amount of time to establish the equilibrium. In contrast to the phase transition region, in which there are a large number of point defects, in the highly ordered region defects in the process of modeling are mainly determined by the boundaries of the domain structure of rhombuses with different orientations.

For these temperatures, at the chemical potential $\mu = 6$, there is a region with lamella residues (LR) in the system, which is ordered at lower temperatures and forms a phase of molten lamellas. As a result, there is no phase transition for this type of structure (Fig. \ref{fig:DenseConc}). In analogy with the ordered phase of rhombuses or bubbles, the times to reach the concentration equilibrium are small and comparable with that for ordered rhombuses. The reason can be that for our system  a strong short range ordering does exist in the disordered phase at not too high temperatures.   

The temperature $T =  0.95$ for this system according to the Ref. \cite{almarza:14:0} is close to critical. The effects of long relaxation time associated with the existence of a short range ordering at this temperature still manifest themselves. Even at a higher temperature $T = 1.2$ in this region, the echoes are observed. The largest time to establish equilibrium is $\tau_{ct} \approx 100$ MCS.

\subsection{Interaction energy evolution}
\label{sec:energy}

\begin{figure}
	\centering
	\includegraphics[width=1\linewidth]{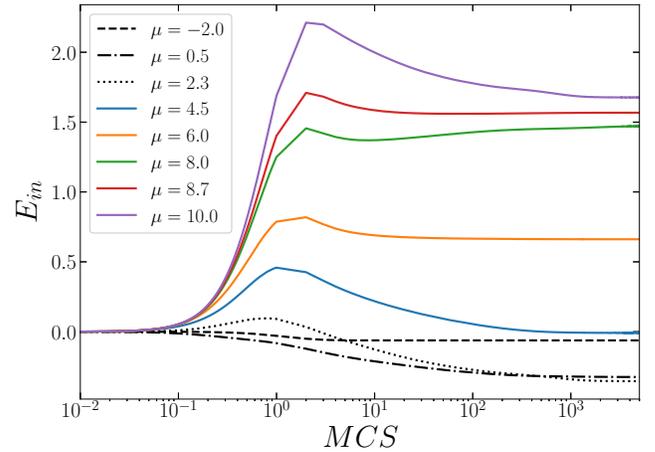}
	\caption{Types of energy relaxation at $T=0.8$ and different chemical potential values }
	\label{fig:typesEnergy}
\end{figure}

Alongside with the concentration evolution, the system energy evolves as well. The concentration manifests the evolution of the one-particle distribution function, while the energy evolution can shad light on the evolution of multiparticle distribution functions. The time evolution of the energy per lattice site 
\begin{equation}
\begin{split}
E_{\rm in}(t) &=\frac {1}{2M}\bigg \langle \sum_{i=1}^{M}\hat {\rho_i} V_1(\mathbf{x},t) \bigg \rangle \\
&=\frac {1}{2M} \bigg \langle \sum_{i=1}^{M}\hat {\rho_i}(-z_{1i} J_1+z_{3i} J_3) \bigg \rangle,
\end{split}
\label{eq:averagenergy}    
\end{equation}
shows more diverse behavior (Fig.\ref{fig:typesEnergy}) as compared to the concentration evolution.

For small values of the chemical potential, the energy decreases monotonically, which is caused by the presence of clusters with negative energy in the system: dimers, triangles, and rhombuses arise under the influence of the attraction of the nearest neighbors.

\begin{figure*}
	\centering
	\includegraphics[width=1\linewidth]{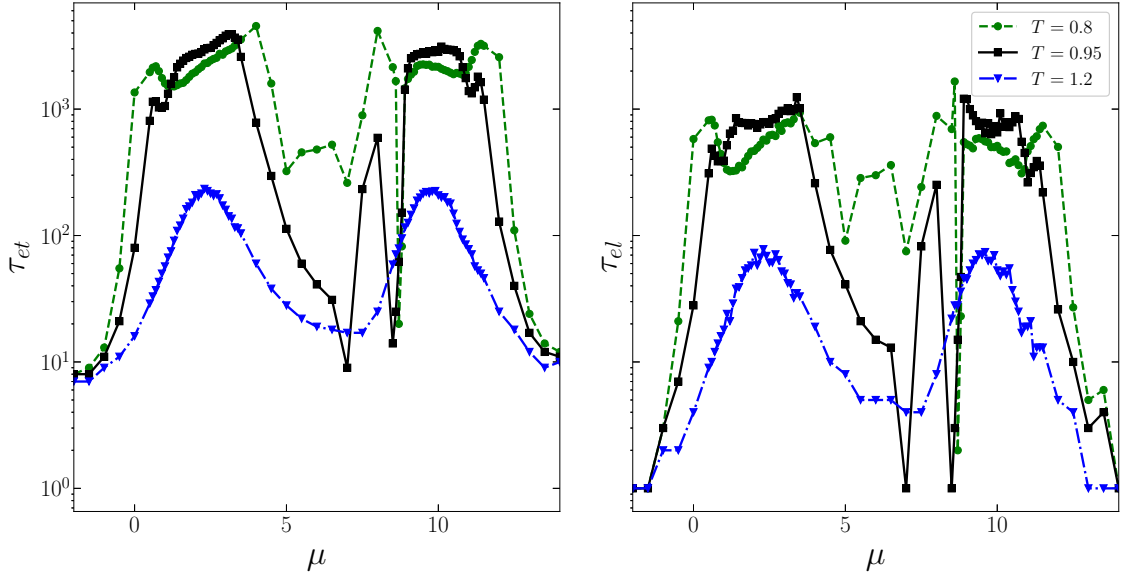}
	\caption{The total (left panel) and longest (right panel) energy relaxation time versus the chemical potential at different temperatures.}
	\label{fig:TimeEnergy}
\end{figure*}

With an increase in the chemical potential, pairs of third neighbors are formed in the system at the beginning that increase the energy of the system, after which rhomboid clusters are formed up to their ordered phase and thereby cause decreasing the energy. Subsequent concentration saturation leads to the fact that the shape of the energy relaxation curve has characteristic maximum and minimum points that reflect the priority of the interaction between particles. (The maxima manifest the preference of repulsion and the minima demonstrate importance of attraction interactions) at different stages of reaching equilibrium. At conditions when the region with lamella residues exists, the competing interaction leads to a longer equilibrization of the multiparticle distribution as compared to the concentration evolution (Fig.\ref{fig:TimeEnergy}, left). As in the case of concentration relaxation, the total energy relaxation time $\tau_{et}$ to reach the equilibrium state for internal energy was determined through the difference between the equilibrium and the current value smoothed out over 11 points. The equilibrium energy value was also determined by averaging over the last $100$ MCS and over $20000$ trajectories. The range of energy change in the system is larger by an order of magnitude as compared to the range for concentration, which is why when the difference $|E_{\rm eq} - \sum_{j = -5}^5 E_{i+j}/11|$ reaches $10^{-3}$, the integral of the energy over time reaches its constant value with the necessary precision. 

In the regions of unstable and disordered phases, reaching equilibrium is accompanied by a long monotonous process, in which the energy tends to an equilibrium value. This stage, as in the case of concentration, was identified through the longest relaxation time, which behaves like the total relaxation time, being several times shorter.   

The distribution of the relaxation times, as well as their absolute values are mainly in correspondence with that for the concentration evolution. The significant differences arise at the chemical potential values $\mu \in (5;7)$ where region with lamella residues exist. The energy total relaxation time in this region is considerably larger of the concentration relaxation time. The mutual redistribution of particles lasts for a long time after the concentration reaches its equilibrium value. A complicated temperature and chemical potential dependence of the total relaxation time in this region can be noted as well. The distribution of the energy relaxation times is not symmetric with respect to the concentration 0.5 due to more complicated particle redistribution in the crowded environment at $c>0.5 (\mu>6.0)$. At temperatures 0.8 and 0.95, a deep minimum of the energy total relaxation time is observed at $\mu \approx 8.7 $ that does not correspond to the highly ordered state of the rhombus bubbles that exists at $\mu=9.6$ and $c=2/3$. In this region, there is a kind of gap: the energy fluctuations over the course of the simulation are small due to the competition between the attractive and repulsive parts of the interaction.

In the system, saturation of particles on the surface occurs within a short time interval in the region of ordered rhombuses ($\mu=2.4$) or bubbles ($\mu=9.6$). In this case, the concentration turns out to be very close to equilibrium, but there is still no true equilibrium state. In a few steps, the particles build up a domain structure consisting of diamond-shaped clusters of different orientations. During further simulation, the clusters choose only one priority direction of orientation (there can be 3 of them in total) and the domain structure disappears. In this case, the energy of the system is less than at the initial saturation. This process takes a significant amount of time, which reflects the difference between concentration and energy relaxation in these regions.

\subsection{Finite size effects}
\label{sec:size}

Although periodic boundary conditions facilitate reduction of size effects, at some conditions (e.g., in the vicinity of phase transition lines or critical points\cite{landau:76:1,landau:83:0,binder:74:0}) size effects can be important. To understand the influence of the system size on the characteristics of adsorption kinetics, the system of size 120x120 lattice sites was considered. As larger systems require larger computational resources, less number of the system state points and averaging over 2 000 against 20 000 MC trajectories were considered. Several series of MC simulations with averaging over 2 000  trajectories at some particular thermodynamic conditions were performed. The variation of the results around 5 and 10 percent for the concentration and energy relaxation times, respectively, were observed. Thus, the precision of simulation after averaging over 20 000 trajectories can be estimated around 2--3 percent.     
The simulation results at the elevated temperature $T=1.2$ show (Table) that the finite size effect is negligible at this temperature. Small increase of the relaxation times at $\mu=1.9 (c\simeq 0.33)$ is only observed in the region of the ordered states at lower temperatures.

\renewcommand{\arraystretch}{1.6} 
\renewcommand{\tabcolsep}{0.5cm}   
\begin{table}[H]
	\caption{The total ($\tau_{ct}$) and longest ($\tau_{cl}$) concentration relaxation times on the lattices of $60\times 60$ and $120\times 120$ lattice sites at $T=1.2$. \label{tab:points}}
	\begin{center}
		\begin{tabular}{|c|c|c|c|c|}
			\hline
			& \multicolumn{2}{c|}{$\tau_{ct}$} & \multicolumn{2}{c|}{$\tau_{cl}$}\\
			\cline{2-5}
			\raisebox{1.5ex}[0cm][0cm]{$\mu$}
			& 60 & 120 & 60 & 120  \\
			\hline
			0.0 & 15 & 15 & 5& 5 \\
			\hline
			1.2 & 79 & 77 & 28& 28 \\
			\hline
			1.5 & 115 & 113 & 47& 40 \\
			\hline
			1.9 & 123 & 149 & 55& 70 \\
			\hline
			2.3 & 42 & 39 & 27& 21 \\
			\hline
			2.4 & 7 & 7 & 4& 3 \\
			\hline
		\end{tabular}
	\end{center}
\end{table}

\begin{figure*}
	\centering
	\includegraphics[width=1\linewidth]{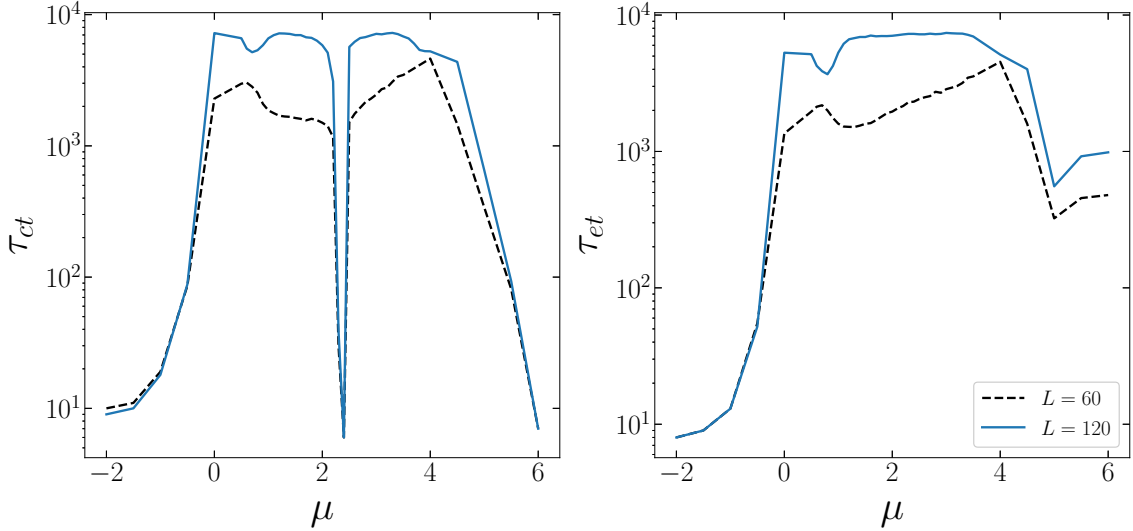}
	\caption{The total concentration $\tau_{ct}$ (left panel) and interaction energy $\tau_{et}$ (right panel) relaxation times on lattices of 60$\times$60 and 120$\times$120 lattice sites for $T=0.8$}.
	\label{fig:sizeeffect}
\end{figure*}

At lower temperature $T=0.8$, the size effect is discernible as it follows from Fig.\ref{fig:sizeeffect}. The relaxation times in the region of the ordered phases or close to them in the enlarged system are 2--3 times larger. It is not a large difference on the background of almost 4 orders variation of the relaxation times. What is more important, the chemical potential (or concentration) dependence of the relaxation times are similar for the systems with $L=60$ and 120. In both cases, we can see very narrow minima for the concentration relaxation at the chemical potentials corresponding to the most ordered system states ($\mu=2.4, c=1/3$ and $\mu=6.0, c=0.5$). 

The total concentration relaxation times for the systems with $L=30$, 60, 120 and 180 and averaging over 2 000 MC trajectories were found equal correspondingly to 2436, 2745, 6562 and 6848 MCS. The relaxation times depend on the system size strongly nonlinear. The most significant increase 2.39 times of the relaxation time reached with increasing the system size from $L=60$ to 120, while it only increased 1.04 times when the system size enlarged from 120 to 180. Probably, the finite size effect will be negligible for systems with $L\geq 200$ at $T=0.8$. 

Similar influence of the system size on the energy relaxation times is observed as well (right panel on the Fig.~\ref{fig:sizeeffect}).

The finite size effect is quite sensitive to the temperature change. Very large systems may be necessary to study adsorption processes at lower temperatures. Investigation of the adsorption finite size effects in cluster forming systems can be a subject of a separate publication.

\section{ Discussion and conclusion}
\label{sec:discussion}

The master equation is used for describing the kinetics of adsorption of particles with competing interaction on a flat surface. The thermally activated adsorption and desorption transition rates are suggested to model the sticking probabilities. It is shown that the inverse value of the frequency prefactor of the transition rates is the time scale for transferring the Monte Carlo steps into physical time. 

The total relaxation time was determined as time when the integral of the difference between the function and its equilibrium value starts to be independent on time. It was observed that this time is that the lattice concentration and interaction energy reach the values differing from their equilibrium values by $10^{-4}$ and $10^{-3}$, respectively. The longest relaxation time was determined as the time interval during which the deviation of the function from its equilibrium value decreases by e times just before time reaches the total relaxation time. The longest relaxation times are several times shorter of the total ones because they characterize last parts of the relaxation curves. However, both times demonstrate similar behavior as functions of the chemical potential and temperature. 

The concentration evolution during the first Monte Carlo step is fast. The concentration reaches values comparable with the equilibrium concentration by the end of the first MCS. Three different types of the subsequent concentration evolution was observed depending on the final equilibrium state of the system. A simple exponential decay of the concentration deviation from the equilibrium value was observed at low or high equilibrium concentration corresponding to disordered gas-like distribution of particles or vacancies in the system. More complicated still monotonic concentration behavior is characteristic for equilibrium concentrations corresponding to ordered rhombuses or rhomboidal bubbles phases. For concentrations at which lamella exist, the overshooting behavior is demonstrated. The concentration on an earlier stage of relaxation attains values larger of the equilibrium ones. 

Alongside with the variety of the relaxation curves shape, the relaxation times span over almost four orders of magnitude. The largest relaxation times are attained at concentrations corresponding to phase transitions between ordered and disordered states. Very narrow minima exist for ideally ordered rhombuses or rhomboidal bubbles at equilibrium concentrations 1/3 or 2/3 ($\mu=2.4$ or 9.6). An additional deep minimum at concentration 1/2 ($\mu=6.0$) can be attributed to the peculiarities of the phase diagram of the system close to this concentration\cite{almarza:14:0}. A narrow region of ordered lamellas was found there at rather low temperatures. However, in our simulation the system was four times smaller ($L=60$ against $L=120$ in Ref.\cite{almarza:14:0}); thus, the critical points can be shifted to higher temperatures with decreasing the system size\cite{landau:76:1,landau:83:0}. 

The short range ordering can exist at temperatures not significantly larger of the critical temperatures. Then the barrier resistance due to creation of ordered structures\cite{litniewski:19:0,bildanau:20:0} strongly hampers the adsorption of particles that leads to large relaxation times at the temperatures 0.8 and 0.95. However, at the concentrations corresponding to the ideal ordering the effect of the system self-organization ensures fast adsorption. With the temperature increase, thermal fluctuations destroy the ordering that leads to decrease of the relaxation times by two orders of magnitude already at $T=1.2$. 

The energy relaxation curves show more complicated behavior due to the competition between attractive and repulsive interactions. The energy relaxation times are comparable with that for concentration evolution except of the region where lamellas and ordered rhombuses (bubles) exist. At these conditions, the particle mutual redistribution lasts several times longer of the concentration relaxation. 
On the other hand, in the regions $\mu = 5, 7, 8.7$, the energy reaches the equilibrium value rather quickly, in contrast to concentration. This effect is caused by the competition of interactions between particles during the simulation: energy fluctuations at small times are comparable to fluctuations at large times, when equilibrium is reached in the system. In theese cases, the real relaxation time should be considered as a combination of relaxation of both characteristics (both concentration and internal energy).

Adsorption modelling in cluster forming systems requires considerable statistics. Averaging over 2 000 MC trajectories provide accuracy of about 5 percent concerning the relaxation times estimation. The finite size effects are negligible for the system of 60$\times$60 lattice sites at above critical temperature $T=1.2$. However, the finite size effects are very sensitive with respect to the temperature change. At lower temperature $T=0.8$ the size effects can be negligible in considerably lager system of 200$\times$200 lattice sites. Understanding the finite size effects in more detail requires additional large scale simulation.

In the current research we choose equal the frequency prefactors for adsorption and desorption transition rates. This allowed us to develop the kinetic Monte Carlo algorithm leading to the equilibrium states of the system equivalent to the results of the grand canonical equilibrium Monte Carlo simulation\cite{almarza:14:0}. It is important to note that other choices will lead to the equilibrium states that depend on the ratio of the prefactors. This means that the final adsorbate equilibrium state depends on the details of the particle exchange between the solution and adsorbed phase. Additional factors requiring investigation are the attraction/repulsion of the adsorbent surface and lateral diffusion of the adsorbate.  

\section{Acknowledgements}
The authors thank Prof. Alina Ciach for careful reading and fruitful discussion of the manuscript. This project has received funding from the European Union’s Horizon 2020 research and innovation program under the Marie Skłodowska-Curie grant agreement No 734276.

\section{Authors contributions}
All the authors were involved in the preparation of the manuscript.
All the authors have read and approved the final manuscript.
%
% BibTeX users please use

\bibliography{bibliography_19}

\end{document}